\documentclass[sigconf]{acmart}
\AtBeginDocument{%
  }

\copyrightyear{2025}
\acmYear{2025}
\setcopyright{rightsretained}
\acmConference[CUI '25]{Proceedings of the 7th ACM Conference on Conversational User Interfaces}{July 8--10, 2025}{Waterloo, ON, Canada}
\acmBooktitle{Proceedings of the 7th ACM Conference on Conversational User Interfaces (CUI '25), July 8--10, 2025, Waterloo, ON, Canada}
\acmDOI{10.1145/3719160.3737633}
\acmISBN{979-8-4007-1527-3/2025/07}

\usepackage{subcaption}

\begin{document}

\title{AInsight: Augmenting Expert Decision-Making with On-the-Fly Insights Grounded in Historical Data}

\author{Mohammad Abolnejadian}
\affiliation{%
  \institution{Cheriton School of Computer Science}
  \institution{University of Waterloo}
  \city{Waterloo}
  \state{Ontario}
  \country{Canada}
}
\email{mabolnej@uwaterloo.ca}
\orcid{0000-0003-2028-9270}

\author{Shakiba Amirshahi}
\affiliation{%
  \institution{Cheriton School of Computer Science}
  \institution{University of Waterloo}
  \city{Waterloo}
  \state{Ontario}
  \country{Canada}
}
\email{s2amirsh@uwaterloo.ca}
\orcid{0009-0009-1524-8643}

\author{Matthew Brehmer}
\affiliation{%
  \institution{Cheriton School of Computer Science}
  \institution{University of Waterloo}
  \city{Waterloo}
  \state{Ontario}
  \country{Canada}
}
\email{mbrehmer@uwaterloo.ca}
\orcid{0000-0001-5524-2291}

\author{Anamaria Crisan}
\orcid{0000-0003-3445-3414}
\affiliation{%
  \institution{Cheriton School of Computer Science}
  \institution{University of Waterloo}
  \city{Waterloo}
  \state{Ontario}
  \country{Canada}
}
\email{amcrisan@uwaterloo.ca}








\renewcommand{\shortauthors}{Abolnejadian et al.}

\begin{abstract}
In decision-making conversations, experts must navigate complex choices and make on-the-spot decisions while engaged in conversation. Although extensive historical data often exists, the real-time nature of these scenarios makes it infeasible for decision-makers to review and leverage relevant information. This raises an interesting question: What if experts could utilize relevant past data in real-time decision-making through insights derived from past data? To explore this, we implemented a conversational user interface, taking doctor-patient interactions as an example use case. Our system continuously listens to the conversation, identifies patient problems and doctor-suggested solutions, and retrieves related data from an embedded dataset, generating concise insights using a pipeline built around a retrieval-based Large Language Model (LLM) agent. We evaluated the prototype by embedding Health Canada datasets into a vector database and conducting simulated studies using sample doctor-patient dialogues, showing effectiveness but also challenges, setting directions for the next steps of our work.
\end{abstract}




\maketitle

\section{Introduction}
\label{sec:intro}

Decision-making is an essential part of daily life, with an even
greater impact in critical areas such as healthcare. Making effective decisions in such domains is demanding even for experts for two key reasons. First, decisions in these areas require a high level of precision because of their importance. Second, experts must consider multiple aspects of the problem, relying heavily on prior knowledge and
historical data from the field. However, retrieving such data and
getting helpful insights in real time can be challenging. Although large-scale data for these domains, such as \cite{NICEGuidance, HealthCanadaPortal} are available, they tend to be underutilized during the decision-making process due to information overload and limited time. For instance, a physician treating a patient with a rare symptom might not be aware of case studies published a week before on such symptom. With the progress of Artificial Intelligence (AI) technology in generative domains, coupled with information retrieval techniques, timely interventions are now a feasible approach that can be utilized to aid the experts in the decision-making process. In this work, we aim to create a conversational user interface that leverages a Large Language Model (LLM) agent supplemented with a Retrieval Augmented Generation (RAG)\cite{Lewis2020-pf} system to provide experts with on-the-fly insights that are extracted from the existing knowledge base in their field, leading them to make more informed choices. Specifically, we will use the retrieval system to retrieve data related to the ongoing decision-making problem and employ the LLM to present this prior knowledge in a succinct way for quick interpretation. As an example of how this system can be used, we specifically focus on an ongoing conversation between a doctor and a patient where the patient is discussing their symptoms while the doctor is making decisions on possible treatment options. This will help us see how our proposed system can work in a real-world scenario where an expert is making a decision while having a conversation. Our approach in this work will flow through three main phases: first, the design and implementation of the core processing pipeline responsible for analyzing the conversation, retrieving relevant past data, and generating timely insights. Second, we create a user interface to present these insights clearly and accessibly to the expert during their decision-making process. Finally, we curate and embed a domain-specific dataset to conduct simulated studies and see the system’s functionality as a preliminary evaluation. Specifically, our system continuously listens to the ongoing conversation in the background. As the conversation goes on, the system generates insights based on the information relevant to the context of the conversation and retrieved from the knowledge base, enabling the expert to further evaluate their decisions and refine their choices as needed.

For the outcome of this work, we expect to find answers to the following research questions:
\\


\noindent \textbf{RQ1:} How can on-the-fly insights elicited from an expert's domain's historical data assist them in making more informed decisions while being involved in a synchronous decision-making situation?

\vspace{1em}

\noindent \textbf{RQ2:} How well can a combination of a retrieval system and a language model generate timely insights?

\vspace{1em}

Our approach aims to improve the reliability and efficiency of expert decisions in high-stakes domains while maintaining human agency in the decision-making process by delegating the final decision to the expert while assigning the AI-based tool the role of providing supplementary information, fostering a more trustworthy human-AI collaboration. In contrast to previous approaches~\cite{Kim2024-mh}, our system aims to assist rather than replace human decision-makers in such domains, allowing them to make the final decisions while leveraging past knowledge. While this work proposes a novel methodology to eventually answer these research questions, further empirical validation through actual user studies conducted in relevant professional domains is necessary to fully assess the method's impact.
\section{Related Work}
\label{sec:related_work}

Decision-making is a demanding task, often leading to decision fatigue as the decision-maker must carefully evaluate multiple aspects of the problem and predict the potential consequences of their choices\cite{Pignatiello2020-do}. This problem is even more pronounced in high-stakes domains such as healthcare, where experts are required to make decisions that have much more impactful consequences\cite {Persson2019-af}. This brings up the need for tools to support the decision-making process, especially for aiding experts in this task. 

With the advent of generative AI technology, there has been a rising desire for AI-assisted decision-making, with works such as \cite{Chiang2024-xm} exploring the use of an LLM as a devil's advocate to complement a group's decision-making process. Cao et al.\cite{Cao2024-wc} also proposed an LLM-assisted approach for public sector decision-making, especially in the context of environmental policy and climate change. 

However, utilizing generative AI in such critical tasks presents different challenges like trust and reliability\cite{Steyvers2024-pc}. Therefore, understanding how experts interact with and trust AI is essential for designing effective AI-assisted decision-making tools. Zhang et al.\cite{Zhang2020-th} examined two AI-assisted decision-making scenarios. The findings show that the system should appropriately calibrate human trust in the AI, revealing that generative AI technologies alone lack sufficient trust to be used in critical decision-making tasks. Recognizing the importance of trust calibration in AI, some recent works have explored advanced retrieval systems to improve the reliability of LLMs in decision-making. Jan et al.\cite{Jang2024-wf} introduced a novel framework called Calibrated  Retrieval-Augmented Generation \textit{(CalibRAG)} that provides LLMs not only with relevant documents but also confidence levels for each document that ensures reliable decision-making. With the reliability of generated answers being even more important in critical domains such as health care, Hammane et al.\cite{Hammane2024-zg} proposed a new approach called \textit {SelfRewardRAG} that incorporates the RAG system with LLMs to receive up-to-date information from medical repositories. Their pipeline consists of a self-evaluation and resolution layer that assesses and refines the LLM responses to ensure the highest accuracy in the medical domain. Kim et al.\cite{Kim2024-mh} take this work further by proposing a more complex network of LLMs working with each other to make decisions in a medical setting and considering a moderator as well by introducing \textit{MDAgents}. The key takeaway of this work is a component named \textit{MedRAG}, which improves \textit{MDAgents}'s average accuracy by $4.7\%$, and when all of this is complemented by a moderator's review (as the expert), the average accuracy of \textit{MDAgents} improves by $11.8\%$.

While trustworthiness and reliability are crucial aspects of AI-assisted systems, their effectiveness also depends on how users interact with them. Conversational User Interfaces (CUI) address this need by providing a natural way of communication, particularly in a decision-making context. Liu et al.\cite{liu_conversational_2024} compared the task completion time and user perception when the users utilize CUIs in comparison with Graphical User Interfaces (GUI) for decision aids. The results demonstrated that neither of them surpasses the other, emphasizing the need for designing UIs tailoring to users need and cognitive styles. However, Gupta et al.\cite{gupta_trust_2022} showed that CUIs receive more user trust in decision support systems compared to conventional GUIs.

Building upon the effectiveness of CUIs and retrieval systems in the decision-making context, our proposed system integrates both CUIs and retrieval systems to assist experts in their decision-making tasks by generating on-the-fly insights based on past relevant data.
\section{Methodology}
\label{sec:methodology}

Building on the motivation and related works, we developed a prototype aimed at supporting real-time expert decision-making by generating on-the-fly insights from historical data, focusing on doctor-patient dialogue as an example use case. Our system integrates a core processing pipeline to analyze conversations and generate insights, a User Interface (UI) to present this information, and an embedded knowledge base derived from relevant datasets. This structure facilitates the simulated evaluations presented in Section~\ref{section:demonstration}.

\subsection{System Overview}
\label{subsec:system_overview}

\begin{figure*}[t]
    \centering
    \includegraphics[width=\textwidth]{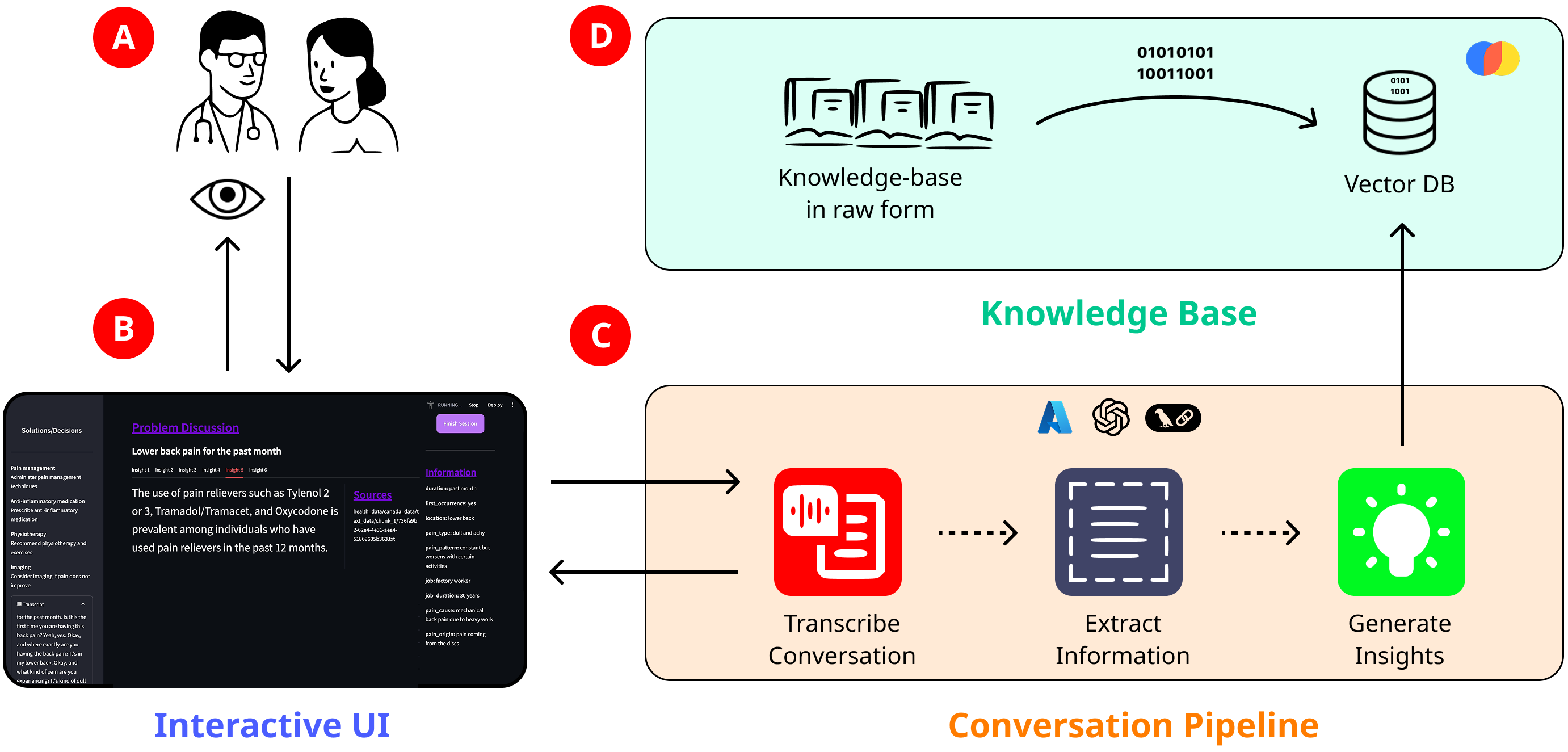}
    \caption{System overview illustrating different components of our prototype and how they interact with each other: (A) Ongoing conversation between the doctor and the patient, (B) Interactive conversational interface displaying real-time insights along with other relevant information from the dialogue (Refer to Section~\ref{sec:pipeline}), (C) Conversation processing pipeline handling transcription, information extraction, and insight generation, and (D) Knowledge base consisting of the decision-maker's relevant historical data, embedded into a vector database.}
    \Description{A diagram illustrating the AInsight system's workflow, divided into four main interconnected components labeled A, B, C, and D, along with three larger conceptual areas: "Interactive UI," "Conversation Pipeline," and "Knowledge Base."
    Component (A) depicts a doctor and a patient in conversation, symbolizing the real-time audio input.
    Component (B) shows a screenshot of the "Interactive UI," a dark-themed interface with four panels: "Solutions/Decisions" (listing "Lower back pain for the past month"), "Problem Discussion" (showing example text "The use of pain relievers such as Tylenol 2 or 3..."), "Sources" (listing a file path), and "Information" (with fields like "duration: month," "first occurrence: past," "location: lower back").
    Component (C) represents the "Conversation Pipeline," which consists of three sequential blocks connected by rightward arrows: "Transcribe Conversation," "Extract Information," and "Generate Insights." This pipeline takes input from the conversation and interacts with the knowledge base.
    Component (D) illustrates the "Knowledge Base." It shows "Knowledge-base in raw form" (represented by a document icon) being transformed (indicated by an arrow with binary code "01010101 10011001") into a "Vector DB" (represented by a database cylinder icon).
    Arrows indicate the flow: The conversation (A) inputs into the Interactive UI (B). The Conversation Pipeline (C) processes this information; its "Generate Insights" step retrieves data from the "Vector DB" within the Knowledge Base (D). The insights and extracted information from the pipeline (C) are then displayed on the Interactive UI (B).}
    \label{fig:system_overview}
\end{figure*}

To provide a high-level understanding of how our system operates, we illustrate the end-to-end flow in Figure~\ref{fig:system_overview}. This overview connects all major components of our system which we will discuss in the following subsections: from the synchronous interaction between the doctor and the patient, to the conversation pipeline consisting of audio transcription, information extraction, and insight generation, and finally to how relevant knowledge is retrieved from a vector database, where the knowledge base was initially embedded and stored. This visualization shows how we envisioned our prototype to be used and how its different components connect and work together. The complete implementation, including source code and detailed setup instructions, is available on our repository at \textbf{\url{https://github.com/ubixgroup/AInsight}}.

\subsection{Conversation Processing Pipeline}
\label{sec:pipeline}

We started by designing a pipeline to orchestrate the entire backend process of our prototype. The pipeline begins with transcribing the audio of the conversation, passing the transcribed text to an LLM to extract the problem and the solutions that are being discussed in the conversation, and finally passing this extracted information to our insight generation module, where data related to this information will be retrieved and given to the LLM model as the context to generate insights. 
This pipeline requires a Machine Learning (ML) model to transcribe the audio recording, an agentic LLM that has access to tools and acts as the brain behind extracting information and generating insights from the retrieved information, an embedding model to embed the whole knowledge base that we collected and later embed the query used to retrieve from the vector database, and finally a framework for sticking all of these pieces together and making a working chain. Specifically, we utilized Microsoft Azure's OpenAI Service\footnote{\url{https://azure.microsoft.com/en-us/products/ai-services/openai-service}}, where we deployed the \textit{Whisper} model for transcription, \textit{GPT-4o} for information extraction and insight generation, and \textit{text-embedding-3-small} for embedding data retrieval purposes. Table \ref{tab:model_deployments} summarizes these deployments and usage metrics from our development and simulated studies. To integrate these components seamlessly, we chose LangChain~\cite{Chase_LangChain_2022} as our primary framework due to its compatibility, ease of use, and flexibility in handling the pipeline's various modules.

\begin{table}[htbp]
  \centering
  \begin{tabular}{|p{2cm}|p{1.5cm}|p{1.7cm}|p{1.6cm}|}
    \hline
    \textbf{Description} & \textbf{Model} & \textbf{Tokens/Min} & \textbf{Cost (USD)} \\
    \hline
    Embedding  & text-embedding-3-small & 40.94 mil. & \$0.82 \\
    \hline
    LLM & GPT-4o & 3.28 mil. & \$9.23 \\
    \hline
     Transcribing & Whisper & 105.7 min. & \$0.76 \\
    \hline
  \end{tabular}
  \caption{Pipeline Models Details and Estimated Usages During Development and Simulated Studies}
  \label{tab:model_deployments}
\end{table}
\vspace{-0.6cm}

Diving deep into the implementation details, each new conversation initializes a conversation agent which is responsible for managing inputs and coordinating interactions between different pipeline modules. 
As of the current implementation of the system, there is a 20-second interval between pipeline invocations; this is a design decision made to achieve higher accuracy by allowing for the use of larger, more powerful language models, which inherently require more processing time.
In the first module, the audio recording of the ongoing conversation is transcribed using an API call to the Whisper model. The resulting text is appended to an cumulative transcript, visible in real-time to the doctor, and subsequently passed to the information extraction module.

The next module has the responsibility to extract key conversational elements, particularly the patient's stated problem, relevant contextual information (e.g., pain characteristics and patterns), and solutions or decisions proposed by the doctor. Since the system is listening to an evolving conversation, this information is not necessarily extracted by each call to the LLM, but rather can be updated as well if new information relevant to the past is presented by either side of the conversation. This extracted or updated information serves as input to the final insight generation module and is displayed alongside other decision-making aids for the doctor's reference.

The final component in our pipeline is the insight-generation module, which leverages the information extracted by the previous module to retrieve the five most relevant documents from the vector database. These documents, combined with the contextual information, are then passed to the LLM so it can produce insights that are directly useful to the ongoing discussion.

An additional complexity addressed by this module is its ability to also work with structured data. LLMs have been shown to perform poorly on structured data (such as tables) \cite{10.1145/3616855.3635752} and since all of the data in a knowledge base is not unstructured text data, our insight generation module should also be able to interpret and work with structured data as well. Recognizing this issue, we used an agentic LLM solution instead of a raw chat completion one, where the LLM has the necessary tools to query the structured data and gain insights from the output of these queries. For this, we leveraged LangChain's Pandas Dataframe Agent
\footnote{\url{https://python.langchain.com/docs/integrations/tools/pandas/}} 
which equips the LLM with the ability to query the structured data using the Pandas framework's commands. Figure \ref{fig:pandas_agent_example} shows an example of how this agent works.

\begin{figure}[H] 
  \centering
  \fbox{
    \parbox{0.9\columnwidth}{
      \textbf{Example of how Pandas Dataframe Agent works:}
      For instance, given a dataframe containing an Age column,
      a prompt such as \textit{"What is the average age of the population?"}
      would trigger the following pandas operation:
      \texttt{df['Age'].mean()}
    }
  }
  \caption{An example of how Pandas Dataframe Agent works}
  \label{fig:pandas_agent_example}
  \Description{A text box illustrating an example of the Pandas Dataframe Agent. The title within the box reads: "Example of how Pandas Dataframe Agent works:". Below this, the explanatory text states: "For instance, given a dataframe containing an Age column, a prompt such as "What is the average age of the population?" would trigger the following pandas operation: df['Age'].mean()". The specific pandas code "df['Age'].mean()" is shown as the resulting operation.}
\end{figure}

All prompts utilized within the pipeline's different stages are detailed in \href{https://github.com/ubixgroup/AInsight}{our repository} under the \texttt{prompts} directory.

\subsection{Conversational Interface Design and Knowledge Base Construction}
\label{subsec:cui_data}

To effectively present the pipeline's output, we proceeded to design an interface for the decision-maker—in this case, the doctor. As previously mentioned, our primary objective while developing the UI was to maximize the amount of useful information displayed while making the interface as minimally distracting as possible, requiring little to no direct manipulation from the user. For this, we implemented a UI using the Streamlit\footnote{\url{https://streamlit.io/}} framework. To handle real-time audio input, which Streamlit does not natively support, we integrated the streamlit-webrtc\footnote{\url{https://github.com/whitphx/streamlit-webrtc}} package. The resulting interface, detailed further in Section~\ref{section:demonstration}, displays essential conversational data (like the ongoing transcript and extracted key points) alongside the generated insights, updating dynamically as the conversation progresses without requiring direct user manipulation.

Finally, in order to run our simulated studies, we needed to collect a knowledge base relevant to our case study. For this, we looked into datasets provided by Health Canada\cite{HealthCanadaPortal}, chosen for their extensive coverage of materials including clinical trials, public health surveys, and regulatory information, ultimately collecting \textbf{2,705} data points from Open Canada API\footnote{\url{https://open.canada.ca/en/access-our-application-programming-interface-api}}. After curating our knowledge base, we needed to make it accessible to our retrieval system for retrieving relevant information in real-time. To accomplish this, we embedded the entire knowledge base using OpenAI's \texttt{text-embedding-3-small} model via Azure's deployment and stored these embeddings in a local \texttt{ChromaDB}\footnote{\url{https://github.com/chroma-core/chroma}} instance. This embedding process transformed our text data into high-dimensional vectors that capture semantic meaning, allowing for efficient similarity searches during retrieval operations in the insight generation module of the pipeline.

\section{System Demonstration}
\label{section:demonstration}

\begin{figure*}[t]
    \centering
    \begin{subfigure}[t]{0.45\textwidth}
        \centering
        \includegraphics[width=\linewidth]{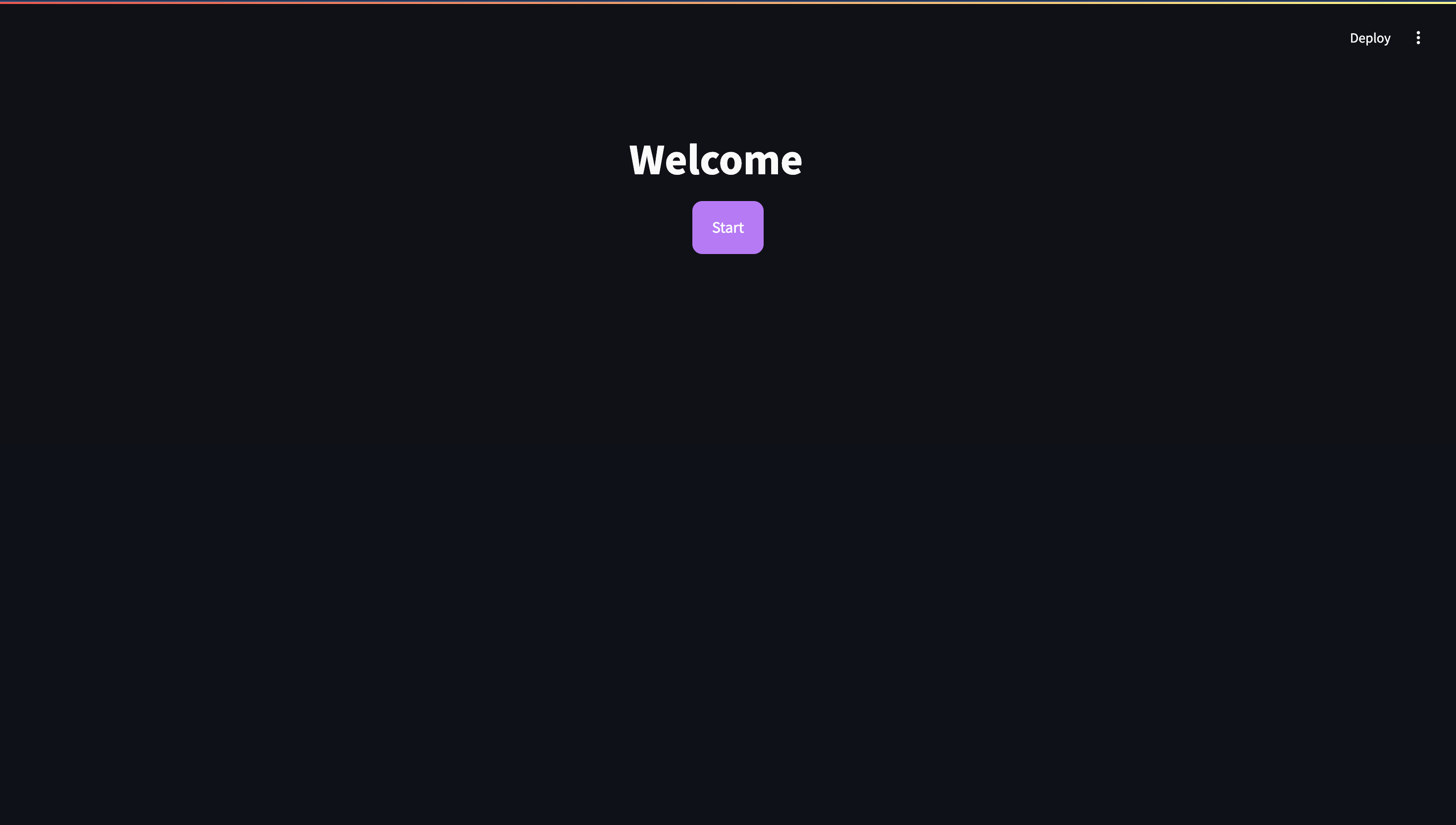}
        \caption{Start page}
        \label{fig:start_page}
    \end{subfigure}
    \hspace{0.05\textwidth}
    \begin{subfigure}[t]{0.45\textwidth}
        \centering
        \includegraphics[width=\linewidth]{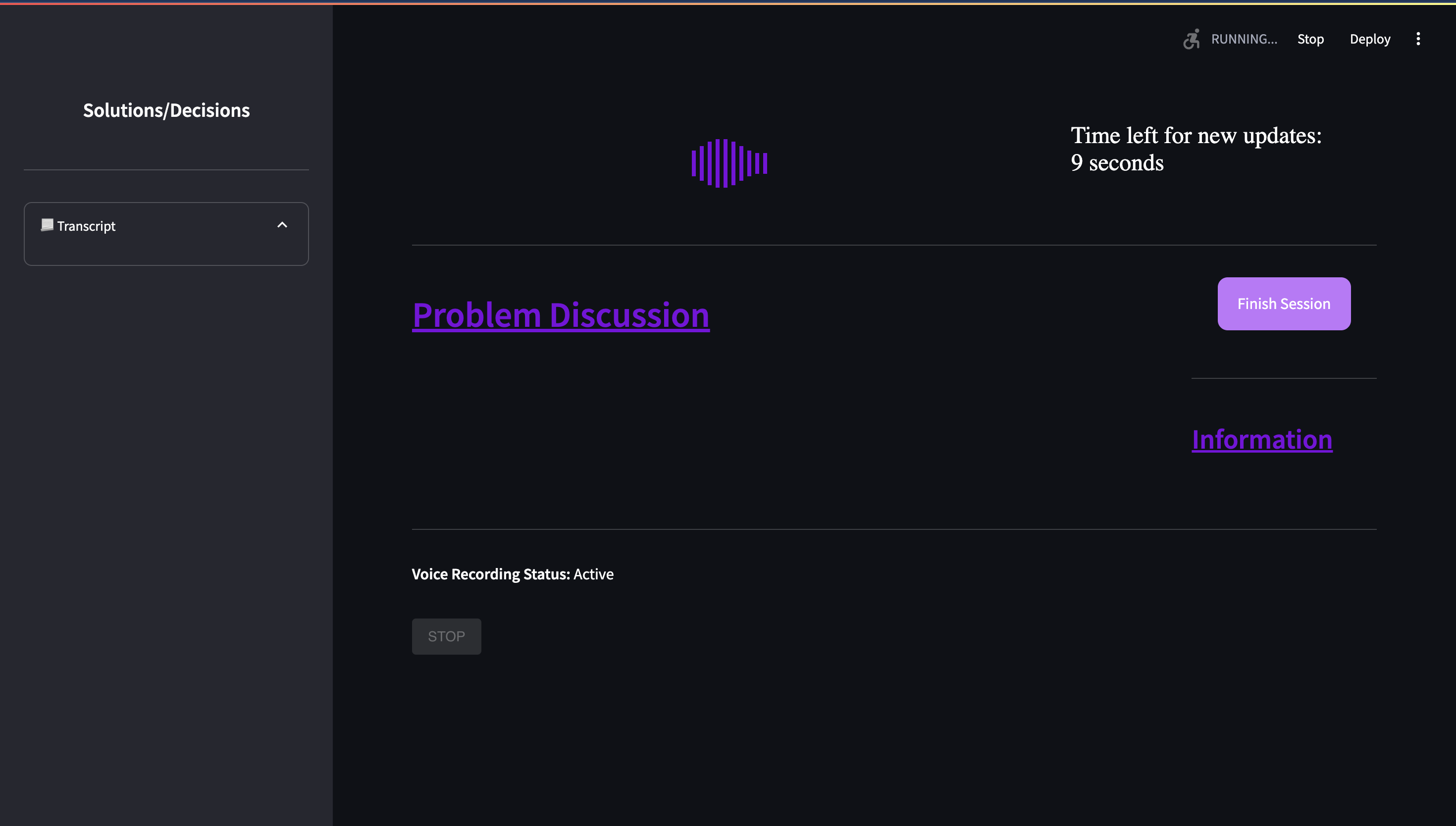}
        \caption{Initial empty interface state}
        \label{fig:voice_record}
    \end{subfigure}

    \vspace{1em} 

    \begin{subfigure}[t]{0.95\textwidth}
        \centering
        \includegraphics[width=\linewidth]{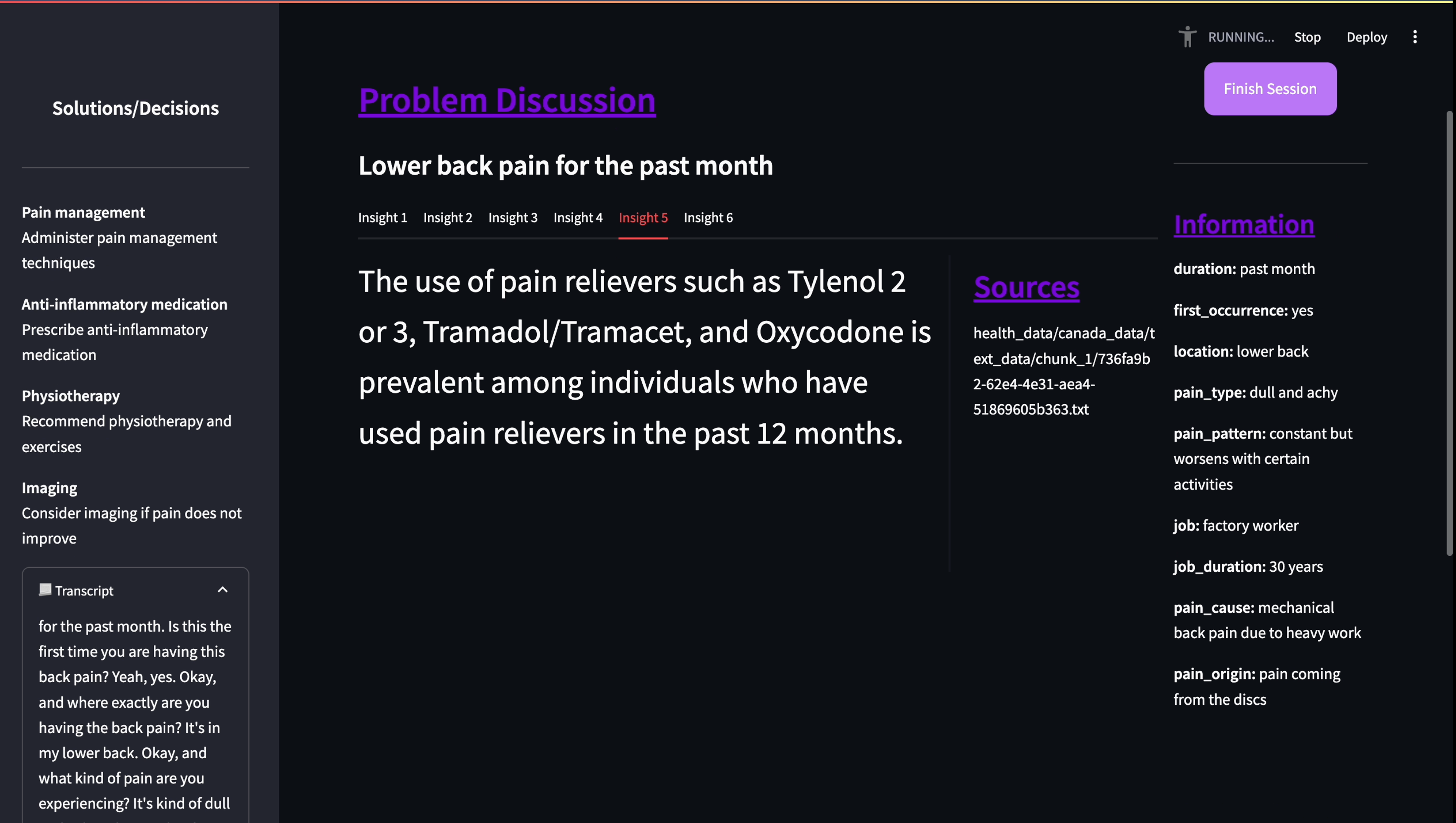} 
        \caption{Interface as the conversation goes on}
        \label{fig:insight_interface}
    \end{subfigure}

    \caption{User interface states demonstrated during an example simulated doctor-patient consultation (transcript on the \href{https://github.com/ubixgroup/AInsight}{repository}): (a) Start page initiating the session, (b) The main interface in its initial empty state after activation, listening to the conversation, (c) The interface populated with extracted information and generated insights as the conversation progressed.}
    \label{fig:insight_illustration}
    \Description{Three screenshots depict the AInsight user interface at different stages of a simulated doctor-patient consultation.
    (a) Start page: A simple dark screen displaying "Welcome" in the center, with a "Start" button directly below it.
    (b) Initial empty interface state: After clicking "Start," the main interface appears with a dark background. It shows a three-column layout. The left column is "Solutions/Decisions" (empty). The middle column is "Problem Discussion" (empty). The right column is "Information" (empty). Below "Problem Discussion," a status indicates "Voice Recording Status: Active" with a "STOP" button. A message "Time left for new updates: 9 seconds" is displayed. At the top right, there's a "RUNNING..." status indicator, "Stop" and "Deploy" buttons, and a "Finish Session" button.
    (c) Interface as the conversation goes on: The interface is now populated with information.
    The "Solutions/Decisions" panel lists items such as "Pain management," "Anti-inflammatory medication," "Physiotherapy," and "Imaging."
    The "Problem Discussion" panel shows "Lower back pain for the past month" and a generated insight: "The use of pain relievers such as Tylenol 2 or 3, Tramadol/Tramacet, and Oxycodone is prevalent among individuals who have used pain relievers in the past 12 months." Below this insight are tabs for "Insight 1" through "Insight 6."
    The "Sources" panel (below the insight) lists a file path: "health_data/canada_data/text_data/chunk_1/736fa9b...".
    The "Information" panel displays extracted details like "duration: past month," "first_occurrence: yes," "location: lower back," "pain_type: dull and achy," "job: factory worker," etc.
    A "Transcript" section (below "Solutions/Decisions") shows a snippet of the conversation: "...for the past month. Is this the first time you are having this back pain?..."
    The "RUNNING..." status and "Finish Session" button remain at the top right.}   
\end{figure*}

\begin{figure*}[t]
    \centering
    \begin{subfigure}[t]{0.3\textwidth}
        \centering
        \includegraphics[width=\linewidth]{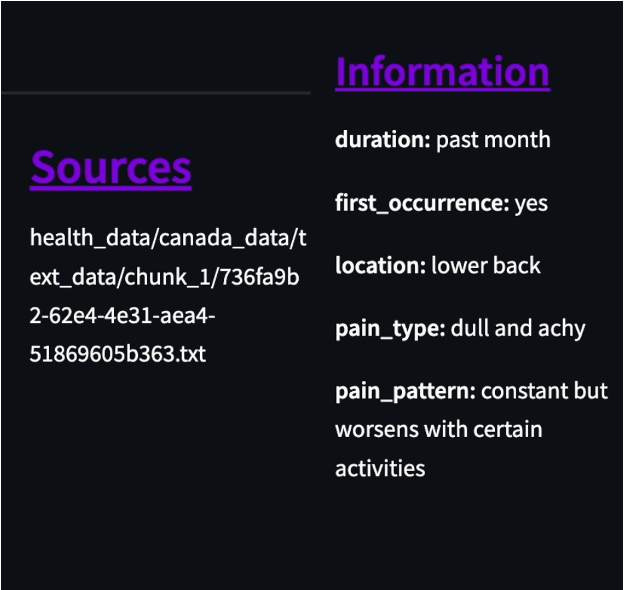}
        \caption{Source and metadata panel}
        \label{fig:source_panel}
    \end{subfigure}
    \hfill
    \begin{subfigure}[t]{0.3\textwidth}
        \centering
        \includegraphics[width=\linewidth]{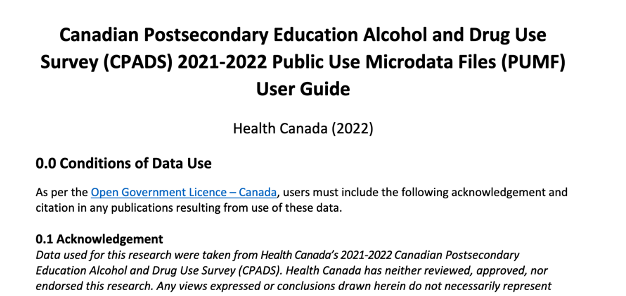}
        \caption{CPADS dataset cover}
        \label{fig:cpads_cover}
    \end{subfigure}
    \hfill
    \begin{subfigure}[t]{0.3\textwidth}
        \centering
        \includegraphics[width=\linewidth]{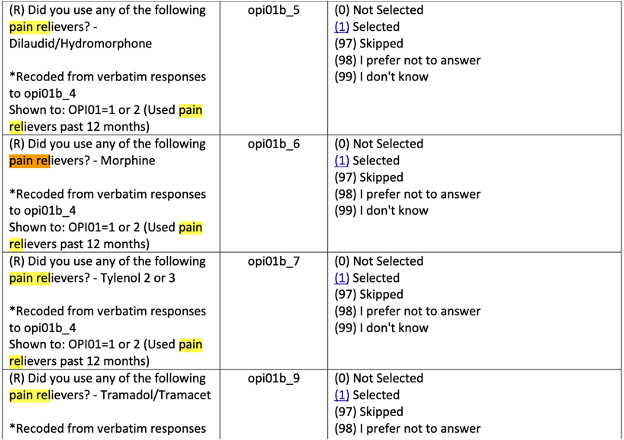}
        \caption{Pain reliever questions}
        \label{fig:pain_relief}
    \end{subfigure}
    \caption{Insight grounding and data traceability example for an insight generated during the example simulation: (a) System UI indicating the data source for an insight, (b) The corresponding CPADS dataset user guide, (c) Relevant survey questions from the guide confirming the insight's basis.}
    \label{fig:insight_validation}
    \Description{Three images demonstrate the traceability of a generated insight back to its source data.
    (a) Source and metadata panel: A cropped screenshot from the AInsight UI. It shows the "Sources" panel listing a file path: "health_data/canada_data/text_data/chunk_1/736fa9b2-62e4-4e31-aea4-51869605b363.txt". To its right, the "Information" panel shows extracted data points: "duration: past month," "first_occurrence: yes," "location: lower back," "pain_type: dull and achy," and "pain_pattern: constant but worsens with certain activities." This panel illustrates where the system displays source information.
    (b) CPADS dataset cover: A screenshot of the cover page of a document. The title reads: "Canadian Postsecondary Education Alcohol and Drug Use Survey (CPADS) 2021-2022 Public Use Microdata Files (PUMF) User Guide." Below the title, it indicates "Health Canada (2022)." Sections "0.0 Conditions of Data Use" and "0.1 Acknowledgement" are visible. This shows the type of source document used.
    (c) Pain reliever questions: A screenshot of a tabular section from the CPADS User Guide. It lists survey questions related to pain reliever use, with variable names and response options. For example, variable "opi01b_7" corresponds to the question "(R) Did you use any of the following pain relievers? - Tylenol 2 or 3" with response options like "(0) Not Selected" and "(1) Selected." Similarly, "opi01b_9" asks about "Tramadol/Tramacet." This image shows the specific questions in the source document that form the basis for an insight about pain reliever usage.}
\end{figure*}

To evaluate the prototype's functionality and test our core idea at this stage of our work, we conducted studies using simulated doctor-patient dialogues ~\cite{Fareez2022}. The figures in this section illustrate one such simulation, the full transcript for which is available on \href{https://github.com/ubixgroup/AInsight}{our repository}. 

During these simulations, the system ran in the background, monitoring the conversation while authors read the dialogues aloud. Upon activating the system via the start button (Figure~\ref{fig:start_page}), the main interface is presented. Initially, the interface is largely empty (Figure~\ref{fig:voice_record}), while the system is listening to the conversation. As the dialogue progresses, the underlying pipeline extracts key conversational elements (as shown in Figure~\ref{fig:insight_interface}): the identified solutions/decisions discussed and the problem stated, dialogue's contextual information (e.g., location of the pain), generated insights relevant to the conversation along with its list of sources from the knowledge base, and the full transcription of the conversation.

Following each simulation, we performed a sanity check and conducted an analysis of the data sources referenced to ensure the generated insights were relevant and grounded in valid parts of the knowledge base, such as tracing information about pain relievers back to specific survey questions in the CPADS user guide, presented in Figure~\ref{fig:insight_validation}. This process verified the system's capability to extract relevant information from the conversation, retrieve related documents from the knowledge base, and finally generate traceable and succinct insights. Furthermore, future user studies could explore presenting metrics such as confidence or similarity scores, similar to those in \cite{Jang2024-yg}, to users in order to evaluate their influence on the decision-making process.
\section{Discussion}
\subsection{Contributions and Design Implications}

In this work, we designed a system to assist experts by providing on-the-fly insights, enabling more informed decisions within limited timeframes. This work contributes to the growing body of research on AI-assisted decision-making through three key contributions. First, our design prioritizes human agency by positioning the system as an augmenting tool, maintaining agency throughout the process, and eventually leaving final decisions to the expert. This emphasis on human control is especially important in domains where decisions carry long-lasting consequences, ensuring experts maintain authority while leveraging AI support. Second, the system emphasizes transparency by grounding generated insights in a knowledge base provided by the expert, fostering trust in the insights and reducing concerns about the unclear origins of LLM-generated responses and occasional hallucinations. Finally, recognizing the real-time constraints of conversational decision-making, our system focuses on presenting supporting information and insights succinctly through a conversational user interface designed for minimal interaction, requiring only navigation between insights

As a work in progress, this work lays the groundwork for enabling more informed decisions in conversation based on trusted knowledge bases. Observations from our simulated studies suggest that our proposed solution and contributions hold strong potential for effective use by experts across various domains, and by addressing challenges and building on future work~\ref{subsec:challenges_future_work}, we believe AI-assisted decision support systems like ours can significantly enhance decision-making processes in various domains, bridging the gap between available information and its real-time application during critical decision moments.

\subsection{Challenges and Future Work}
\label{subsec:challenges_future_work}
During the evaluation of our system, we encountered some challenges, with the first set being concerned with the quality of the generated insights and their backing knowledge base. We found that the effectiveness of the system is largely dependent on the relevance of the indexed data and if the collected knowledge base consists of noisy and irrelevant data, the system may start generating misleading insights which can negatively affect the decision-making process. Therefore, it is important to have high-quality, domain-relevant knowledge bases while ensuring the insight generation module is robust enough to handle potential inconsistencies or noise within the source data. Furthermore, we also encountered challenges regarding the presentation of information and insights. We observed during our simulated studies that even though concise, it still can be distracting for the user to look at the newly generated insights while following the conversation.

Building on these challenges, there are several paths we aim to pursue to extend and refine this work. Primarily, we are looking to conduct comprehensive user studies with experts in a specific domain to further analyze our real-time insight generation system. Through this, we will focus on assessing the depth and perceived usefulness of the insights by the actual end users. Getting feedback from the users and making the pipeline more robust so that it can be more useful, we are also simultaneously working on improving the real-time aspect of our system, looking for ways to minimize user distraction while simultaneously presenting more informative content with clear connections to specific parts of the conversation. 
\section{Conclusion}

In this work, we presented AInsight, a novel approach supporting real-time decision-making through on-the-fly generated insights grounded in the user's existing knowledge base. Our work addresses the challenge faced by experts who must make timely decisions while lacking the ability to fully leverage the extensive historical data available to them. Our prototype implementation, focused on doctor-patient interactions, demonstrates how a system that leverages retrieval-augmented generation can continuously monitor a conversation around a decision-making process, extract key information, and provide contextually relevant insights without requiring direct interaction from the decision-maker. By embedding a comprehensive knowledge base of Health Canada datasets and implementing a pipeline for processing conversational data, we showed that it is possible to surface valuable information that might otherwise remain unused during critical decision points. Our simulated studies revealed both the potential and limitations of our approach. The system successfully retrieved relevant information from the knowledge base. However, challenges remain that motivate the next steps of our work, which include comprehensive user studies and improved ways of presenting generated insights.


\bibliographystyle{ACM-Reference-Format}
\bibliography{sample-base}

\appendix









\end{document}